\documentstyle[aps,pra,graphicx]{revtex}

\draft

\newcommand{\etal}{\textit{et~al.}}

\begin{document}

\twocolumn[\hsize\textwidth\columnwidth\hsize\csname
@twocolumnfalse\endcsname
\title{Collapse-revivals and population trapping in the $m$-photon mazer}

\author{T. Bastin$^{1}$ and E. Solano$^{2,3}$}
\address{$^{1}$Institut de Physique Nucl\'eaire Exp\'erimentale, Universit\'e
  de Li\`ege au Sart Tilman, B\^at. B15, B - 4000 Li\`ege, Belgique\\
$^{2}$Instituto de F\'{\i}sica, Universidade Federal do Rio de
Janeiro, Caixa Postal 68528, 21945-970 Rio de Janeiro, RJ,
Brazil\\ $^{3}$Secci\'{o}n F\'{\i}sica, Departamento de Ciencias,
Pontificia Universidad Cat\'{o}lica del Per\'{u}, Apartado 1761,
Lima, Peru}

\date{\today}

\maketitle

\begin{abstract}
    We present a study of collapse-revival patterns that appear in the changes of atomic
    populations induced by the interaction of ultracold two-level atoms
    with electromagnetic cavities in resonance with an $m$-photon
    transition of the atoms ($m$-photon mazer).
    In particular, sech$^2$ and gaussian cavity mode profiles are considered and differences in
    the collapse-revival patterns are reported. The quantum theory of the $m$-photon mazer is written in the framework of
    the dressed-state coordinate formalism. Simple expressions for the atomic populations, the cavity photon statistics, and the reflection and transmission
    probabilities are given for any initial state of the atom-field system. Evidence for the population trapping phenomenon
    which suppress the collapse-revivals in the $m$-photon mazer is given.
\end{abstract}


\vskip2pc]

\section{Introduction}

    Laser cooling of atoms is a rapidly developing field in
    quantum optics. Cold and ultracold atoms introduce new
    regimes in atomic physics often not considered in the past. Recently, Scully \etal \cite{Scu96} have
    shown that a new kind of induced emission occurs when a micromaser
    is pumped by ultracold atoms, requiring a quantum-mechanical treatment of the center-of-mass
    motion. They called this particular process mazer action to insist on the quantized $z$-motion feature of the induced
    emission.

    The detailed quantum theory of the mazer has been presented in a series of three
    papers by Scully and co-workers \cite{Mey97,Lof97,Sch97}.
    They showed that
    the induced emission probability is strongly dependent on the cavity mode
    profile. Analytical calculations were presented for the mesa and the sech$^2$ mode profiles.
    For sinusoidal modes, WKB solutions were detailed.

    Retamal \etal~\cite{Ret98} showed that we must go beyond the
    WKB solutions for the sinusoidal mode case when we consider
    strictly the ultracold regime. Remarkably, they showed that
    the resonances in the emission probability are not completely
    smeared out for actual interaction and cavity parameters.
    In a recent work \cite{Bas99}, we proposed
    a numerical method for calculating efficiently the induced emission
    probability for arbitrary cavity field modes.
    In particular, the gaussian potential was considered,
    thinking in open cavities in the microwave or optical field
    regime. Differences with respect to the sech$^2$ mode case were found. Calculations for
    sinusoidal potentials were also performed and divergences
    with WKB results were reported, confirming results given in \cite{Ret98}.

    Zhang \etal \cite{Zha99} extended the
    concept of the mazer to the two-photon process by proposing the idea of the two-photon mazer. Their work was
    focused on the study of its induced emission probability in
    the special case of the mesa mode function. Under the
    condition of an initial coherent field state, they showed that this probability exhibits with respect to the interaction length the collapse-revivals
    phenomenon, which have different
    features in different regimes. They are similar to those in the two-photon Jaynes-Cummings
    model only in the thermal-atom regime.

    The collapse-revivals of the atomic excitation in the framework of the
    Jaynes-Cummings model was predicted in the early 1980s by Eberly and co-workers \cite{Ebe80,Nar81,Yoo81}.
    Fleischhauer and Schleich \cite{Fle93} showed later that the shape of each revival is a
    direct reflection of the shape of the initial photon-number distribution
    $P_n$, assuming that the atom is prepared completely in the upper state
    or in the lower state and that the distribution $P_n$ is
    sufficiently smooth. It was also noticed that, under some
    special conditions of the initial atom-field state, the
    revivals can be largely and even completely suppressed \cite{Zah89,Cir90,Gea91}. This phenomenon
    was denominated ``population trapping'' to refer, as noted by Yoo and Eberly \cite{Yoo85}, to a persistent
    probability of finding the atom in a given level in spite of the
    existence of both the radiation field and allowed transitions to other
    levels. The initial atom-field states giving rise to this
    phenomenon were called ``trapping states'' in
    \cite{Cir91}. Let us mention that this denomination is actually used in
    various physical contexts whenever a degree of freedom is found unaltered in spite of the existence of an
    interaction able to change its value. For instance trapping states in the context of the micromaser
    theory have been predicted and very recently measured by Filipowicz \etal \cite{Fil86} and Weidinger \etal \cite{Wei99}
    respectively. Nevertheless, these trapping states
    do not relate with the suppression of the collapse-revivals we are dealing
    here.

    An elegant explanation of the population trapping phenomenon
    has been proposed just very recently by Jonathan
    \etal \cite{Jon99}, who
    noticed that the key to understand the
    collapse-revival patterns under very general conditions is to
    consider the joint initial properties of the atom-field system, even
    if this one is completely disentangled before the interaction. By defining an appropriate coordinate system,
    the dressed-state coordinates, they were able to
    yield simple analytical expressions for the atomic populations which exhibit the conditions needed for population trapping.

    At the present time, no work has been devoted to know
    whether the revivals predicted by Zhang \etal \cite{Zha99} for
    the two-photon mazer may also be suppressed by use of an appropriate
    initial state of the atom-field system. An answer to this question is given
    at the end of this paper. To be not restricted to the case of the two-photon mazer, our analysis
    is generalized to the arbitrary $m$-photon mazer system,
    although the construction of real multiphoton cavities
    results in a formidable experimental task.

    In Sec.~\ref{TheModelSection}, we
    write the quantum theory of the $m$-photon mazer by use of the
    dressed-state coordinate formalism as it was very
    efficient in the description of the population trapping phenomenon in the
    Jaynes-Cummings model~\cite{Jon99}. General expressions are derived for the
    atomic populations and the cavity photon distribution
    after the interaction of the atom with the cavity. The theory is written for any initial pure state of the atom-field system
    (entangled or not). We consider zero temperature and no dissipation in the high-$Q$ cavity. In Sec.~\ref{CollapseRevivalsSection}, results of Zhang \etal \cite{Zha99} are extended to the 1, 2 and 3-photon mazer systems and to various cavity mode
    profiles (mesa, sech$^2$, and gaussian ones). Collapse-revival patterns
    are described for atoms prepared completely in the upper state or in the lower state.
    Sec.~\ref{PopulationTrappingSection} is devoted to the study of the $m$-photon mazer trapping
    states which suppress the collapse-revivals. A brief summary of our results is given in
    Sec.~\ref{SummarySection}.

\section{The model}
    \label{TheModelSection}
    \subsection{The Hamiltonian}
    We consider a two-level atom moving along the $z$-direction in the way to a cavity of
    length $L$. The atom is coupled resonantly with an $m$-photon transition to a single
    mode of the quantized field present in the cavity.
    The atom-field interaction is modulated by the cavity field mode
    function. The atomic center-of-mass motion is described quantum
    mechanically and the rotating-wave approximation is made.
    In the interaction picture, the Hamiltonian describing the system
    is
    \begin{equation}
    \label{Hamiltonian}
        H = \frac{p^2}{2M}+ \hbar g \, u(z) (a^{\dagger m} \sigma + a^m
        \sigma^{\dagger}),
    \end{equation}
    where $p$ is the atomic center-of-mass momentum along the $z$-axis, $M$ is the atomic mass, $\sigma = |b \rangle \langle a|$
    ($|a\rangle$ and $|b\rangle$ are respectively the upper and lower levels of the $m$-photon transition), $a$ and $a^{\dagger}$ are respectively the annihilation
    and creation operators of the cavity radiation field, $g$ is the atom-field coupling strength (half the Rabi frequency) and
    $u(z)$ is the cavity field mode.

    \subsection{The wavefunctions}
    In the $z$-representation and in the dressed-state basis
    \begin{equation}
    \label{DefDressedStateBasis}
        \left \{
        \begin{array}{l}
            |b,0 \rangle, \ldots, |b,m-1 \rangle , \\
            |\pm,n \rangle = \frac{1}{\sqrt{2}}\left(|a,n \rangle \pm |b,n+m
            \rangle\right),
        \end{array}
        \right.
    \end{equation}
    $|n \rangle$ being the photon-number states, the problem reduces to the scattering of the atom upon the
    potentials $V_n^{\pm}(z) = \pm \hbar g \sqrt{(n+1)\ldots(n+m)} \,
    u(z)$. Indeed, the set of wavefunction components
    \begin{equation}
    \label{DefComponents}
        \psi_n^{\pm}(z,t)= \langle z,\pm,n | \psi(t) \rangle ,
    \end{equation}
    where $|\psi(t)\rangle$ is the atom-field state satisfy the
    Schr\"{o}dinger equation
    \begin{equation}
    \label{TimeDependentSchrodingerEquation}
       i \hbar \frac{\partial}{\partial t} \psi_n^{\pm}(z,t) =
            \left( - \frac{\hbar^2}{2M}\frac{\partial^2}{\partial z^2} + V_n^{\pm}(z)\right)
            \psi_n^{\pm}(z,t) .
    \end{equation}

    The general solution of (\ref{TimeDependentSchrodingerEquation}) is
    \begin{equation}
    \label{GeneralSolution}
        \psi_n^{\pm}(z,t) = \int \! dk \, \phi_n^{\pm}(k) e^{-i \frac{\hbar
        k^2}{2M} t} \varphi_n^{\pm}(k,z) ,
    \end{equation}
    where $\varphi_n^{\pm}(k,z)$ is solution of the time-independent
    Schr\"{o}dinger equation
    \begin{equation}
        \label{TimeIndependentSchrodingerEquation}
        \left( \frac{\partial^2}{\partial z^2} + k^2 \mp \kappa_n^2
        u(z) \right)  \varphi_n^{\pm}(k,z) = 0 ,
    \end{equation}
    with
    \begin{equation}
        \label{kappanDef}
        \kappa_n = \kappa \sqrt[4]{(n+1)\ldots(n+m)}
    \end{equation}
    and
    \begin{equation}
        \label{kappaDef}
        \kappa = \sqrt{2M g/\hbar} \: .
    \end{equation}

    The wavefunction components ($n = 1, \ldots, m$)
    \begin{equation}
        \label{psimnzt}
        \psi_{-n}(z,t) = \langle z, b, m - n | \psi(t) \rangle
    \end{equation}
    satisfy a Schr\"{o}dinger equation characterized with a null
    potential and are therefore not affected by the interaction of
    the atom with the cavity. The atom in the lower
    state cannot obviously interact with the cavity field that contains less
    than $m$ photons. The components (\ref{psimnzt}) describe a free particle problem.

    We assume that, initially, the atomic center-of-mass motion is not correlated to the other degrees of freedom.
    We describe it by the wave packet
    \begin{equation}
    \label{InitialStateZ}
    \chi(z) \equiv \langle z | \chi \rangle = \int \! dk A(k) e^{ikz}\theta(-z) ,
    \end{equation}
    where $\theta(z)$ is the Heaviside step function
    (indicating that the atoms are incident from the left of the
    cavity). No restrictions are made for the initial conditions
    of the atomic internal state and the cavity field state, except that pure states are
    only considered. By use of an expansion over the dressed-state basis (\ref{DefDressedStateBasis}),
    we may write
    \begin{equation}
        \label{psi0preamble}
        | \psi(0) \rangle = | \chi \rangle \otimes \left( \sum_{n = 1}^{m} w_{-n}e^{i\chi_{-n}}|b,m - n \rangle +
        \sum_{n = 0}^{\infty} w_{n}e^{i\chi_{n}} | \beta_n \rangle
        \right) ,
    \end{equation}
    with
    \begin{equation}
        | \beta_n \rangle = \cos\left(\frac{\theta_n}{2}\right)|+,n\rangle +
        e^{-i\phi_n}\sin\left(\frac{\theta_n}{2}\right)|-,n\rangle .
    \end{equation}

    The parameters $w_n \in [0,1]$, $\theta_n \in [0,\pi]$ and
    $\chi_n$, $\phi_n \in [0,2\pi]$ are called dressed-state
    coordinates \cite{Jon99}. The normalisation condition is
    \begin{equation}
        \sum_{n=-m}^{\infty}w_n^2 = 1
    \end{equation}
    and the phase factor $\chi_{-m}$ may be set to $0$ without
    loss of generality.

    We consider therefore
    \begin{equation}
        \left \{
        \begin{array}{l}
            \psi_{-n}(z,0) = c_{-n} \chi(z) , \\
            \psi_n^{\pm}(z,0) = c^{\pm}_{n} \chi(z) ,
        \end{array}
        \right.
    \end{equation}
    with
    \begin{equation}
        \left \{
        \begin{array}{l}
        c_{-n} = w_{-n} e^{i \chi_{-n}} ,\\
        c^{+}_n = w_n e^{i \chi_n} \cos\left(\theta_n/2\right) ,\\
        c^{-}_n = w_n e^{i (\chi_n - \phi_n)}
        \sin\left(\theta_n/2\right) .
        \end{array}
        \right.
    \end{equation}

    Inserting Eqs.~(\ref{DefDressedStateBasis}) and (\ref{InitialStateZ}) into Eq.~(\ref{psi0preamble}), we get
    \begin{eqnarray}
    \label{psi0}
    |\psi(0) \rangle & = & \int \!\! dz \! \int \!\! dk \, A(k) \times \nonumber \\
    & & \Bigg( \sum_{n=0}^{\infty} \Big[ S_{a,n} e^{ikz}\theta(-z) |z,a,n \rangle \nonumber \\
    & & + S_{b,n + m} e^{ikz}\theta(-z) |z,b,n+m \rangle \Big] \nonumber \\
    & & + \sum_{n=1}^{m} w_{-n} e^{i\chi_{-n}} e^{ikz}\theta(-z) |z,b,m - n \rangle \Bigg) ,
    \end{eqnarray}
    with
    \begin{equation}
    \label{SanSbn}
            \left(
            \begin{array}{c}
                S_{a,n} \\
                S_{b,n+m}
                \end{array}
            \right)
        = \tilde{A}_n
            \left(
                \begin{array}{c}
                    1 \\
                    1
                \end{array}
            \right)
    \end{equation}
    and
    \begin{equation}
    \tilde{A}_n = \frac{w_n e^{i\chi_n}}{\sqrt{2}}
        \left(
            \begin{array}{cc}
                \cos\left(\theta_n/2\right) & e^{-i\phi_n} \sin\left(\theta_n/2\right) ,\\
                \cos\left(\theta_n/2\right) & -e^{-i\phi_n} \sin\left(\theta_n/2\right) ,
            \end{array}
        \right)
    \end{equation}

    After the atom has left the interaction region, the
    wavefunctions $\varphi_n^{\pm}(k,z)$ can be written as
    \begin{equation}
        \label{varphi}
        \varphi_n^{\pm}(k,z) = \left \{ \begin{array}{ll} r_n^{\pm}(k)e^{-ikz} & (z < 0) \\
                                                          t_n^{\pm}(k)e^{ik(z-L)} & (z > L)
                                                          \end{array}\right. ,
    \end{equation}
    where $r_n^{\pm}(k)$ and $t_n^{\pm}(k)$ are respectively the
    reflection and transmission coefficient associated with the
    scattering of the particle of momentum $\hbar k$ upon the potential $V_n^{\pm}(z)$ (Eq.~\ref{TimeIndependentSchrodingerEquation}).
    The initial state components $\psi_n^{\pm}(z,0)$ have evolved into
    \begin{eqnarray}
        \psi_n^{\pm}(z,t) & = & c^{\pm}_n \int \! dk \, A(k) e^{-i \frac{\hbar
        k^2}{2M} t} \big[r_n^{\pm}(k)e^{-ikz}\theta(-z) \nonumber \\
        & & + t_n^{\pm}(k)e^{ik(z - L)}\theta(z - L) \big]
    \end{eqnarray}
    whereas the free particle wavefunction components $\psi_{-n}(z,0)$ become
    \begin{equation}
        \psi_{-n}(z,t) = c_{-n} \int \! dk \, A(k) e^{-i \frac{\hbar
        k^2}{2M} t} e^{ik(z - L)} \theta(z - L) .
    \end{equation}

    We thus obtain
    \begin{eqnarray}
    \label{psit}
    |\psi(t) \rangle & = & \int \!\! dz \! \int \!\! dk \, A(k) e^{-i \frac{\hbar
        k^2}{2M} t} \times \nonumber \\
    & & \Bigg( \sum_{n=0}^{\infty} \Big[ R_{a,n}(k) e^{-ikz} \theta(-z)   |z,a,n \rangle \nonumber \\
    & &     + T_{a,n}(k) e^{ik(z-L)} \theta(z-L)   |z,a,n \rangle \\
    & &     + R_{b,n+m}(k) e^{-ikz} \theta(-z) |z,b,n+m \rangle \nonumber \\
    & &     + T_{b,n+m}(k) e^{ik(z-L)} \theta(z-L) |z,b,n+m \rangle \Big] \nonumber \\
    & & + \sum_{n=1}^{m} w_{-n} e^{i\chi_{-n}} e^{ik(z-L)} \theta(z-L) |z,b,m - n \rangle \Bigg) , \nonumber
    \end{eqnarray}
    in which
    \begin{mathletters}
    \label{RanRbnTanTbn}
    \begin{equation}
    \label{RanRbn}
        \left(
        \begin{array}{c}
            R_{a,n}(k) \\
            R_{b,n+m}(k)
            \end{array}
        \right)
    = \tilde{A}_n
        \left(
            \begin{array}{c}
                r_n^+(k) \\
                r_n^-(k)
            \end{array}
        \right) ,
    \end{equation}
    \begin{equation}
    \label{TanTbn}
        \left(
        \begin{array}{c}
            T_{a,n}(k) \\
            T_{b,n+m}(k)
            \end{array}
        \right)
    = \tilde{A}_n
        \left(
            \begin{array}{c}
                t_n^+(k) \\
                t_n^-(k)
            \end{array}
        \right) .
    \end{equation}
    \end{mathletters}

    If initially the electromagnetic field is in the state $|n \rangle$ and the atom is in the excited state~$|a \rangle$,
    the only non-zero dressed-state coordinates are $w_n = 1$ and $\theta_{n} = \pi / 2 $. We
    get therefore
    \begin{equation}
        \tilde{A}_n = \frac{1}{2}
        \left(
            \begin{array}{cc}
                1 & 1 \\
                1 & -1
            \end{array}
        \right)
    \end{equation}
    and Eqs.~(\ref{RanRbnTanTbn}) lead to the same results given by Meyer
    \etal \cite{Mey97} who considered in detail this case for the one-photon mazer.

    \subsection{Atomic populations}

    The reduced density matrix $\sigma(t)$ for the atomic internal degree
    of freedom is given by the trace over the radiation and the atomic external variables of the
    atom-field density matrix, that is its elements $i,j = a,b$ are
    \begin{equation}
        \label{sigmat}
        \sigma_{ij}(t) = \sum_{n} \int \!\! dz \langle z,i,n |
        \psi(t) \rangle \langle \psi(t) | z,j,n \rangle .
    \end{equation}

    The atomic populations $\sigma_{ii}$ follows immediately
    from Eq.~({\ref{sigmat})~:
    \begin{equation}
        \label{sigmaiit}
        \sigma_{ii}(t) = \sum_{n} \int \!\! dz |\langle z,i,n |
        \psi(t) \rangle |^2 .
    \end{equation}

    Inserting Eqs.~(\ref{psi0}) and (\ref{psit}) into Eq.~(\ref{sigmaiit}) and using Eqs.~(\ref{SanSbn}) and (\ref{RanRbnTanTbn}), we
    get for an incident atom of momentum $\hbar k$~:
    \begin{eqnarray}
        \label{sigmaAAInitial}
        \sigma_{aa}(0) & = & \frac{1}{2}\left[ 1 - \sum_{n=1}^{m} w_{-n}^2 + \sum_{n=0}^{\infty}
         w_n^2 \sin(\theta_n) \cos(\phi_n) \right] ,\\
        \label{sigmaAAt}
         \sigma_{aa}(t) & = & \frac{1}{2}\left[ 1 - \sum_{n=1}^{m} w_{-n}^2 + \sum_{n=0}^{\infty}
         w_n^2 \sin(\theta_n) \textrm{Re}(e^{i\phi_n} K_n) \right] ,\nonumber \\
    \end{eqnarray}
    where
    \begin{equation}
        \label{defKn}
        K_n = r^+_n r^{-*}_n + t^+_n t^{-*}_n .
    \end{equation}

    The change of the atomic population $\sigma_{aa}$ induced by the interaction of the
    incident atom with the cavity radiation
    field is then given by
    \begin{equation}
        \delta \sigma_{aa} = \sigma_{aa}(t) - \sigma_{aa}(0) ,
    \end{equation}
    with the time $t$ chosen long after the interaction.

    Thus we have
    \begin{equation}
        \label{deltaSigmaAA}
        \delta \sigma_{aa} = \sum_{n=0}^{\infty} \Delta_n ,
    \end{equation}
    with
    \begin{equation}
    \label{deltaN}
    \Delta_n = \frac{w_n^2}{2} \sin(\theta_n) \left[
        \textrm{Re} \left( e^{i\phi_n} K_n \right) - \cos(\phi_n)
        \right].
    \end{equation}

    As expected, the components $w_{-n}e^{i\chi_{-n}}$ of the initial state $|\psi(0)\rangle$ over the states $|b,n
    \rangle$ $(n < m)$ do not play any role in the dynamics of the system.

    We have to emphasize that in Eq.~(\ref{deltaSigmaAA})
    $\Delta_n$ \emph{cannot} be interpreted strictly as the change in the
    $\sigma_{aa}$ population induced by the interaction of the
    two-level atom with the cavity radiation field containing
    $n$ photons. This is only true when the incident atom is prepared in the excited state. Indeed, if initially the internal atomic state
    is $c_a |a\rangle + c_b |b\rangle$ and the field state is
    $|n\rangle$ $(n \geq m)$, then the only non-zero dressed-state coordinates
    are $w_n = |c_a|$, $\chi_n = \arg(c_a)$, $\theta_n = \pi/2$, $w_{n-m} = |c_b|$, $\chi_{n-m} = \arg(c_b)$,
    $\theta_{n-m} = \pi/2$ and $\phi_{n-m} = \pi$. We thus have in
    that case
    \begin{eqnarray}
        \delta \sigma_{aa}  & = & \Delta_n + \Delta_{n-m} ,\nonumber \\
                            & = & \Delta_n \quad \textrm{iff} \quad c_b=0.
    \end{eqnarray}

    \subsection{Photon statistics}

    The reduced density matrix $\rho(t)$ for the cavity radiation
    field is given by the trace
    over the internal and external atomic degrees of freedom of the
    atom-field density matrix, that is its elements $n,n'$ are
    \begin{equation}
        \label{rhot}
        \rho_{nn'}(t) = \sum_{i = a,b} \int \!\! dz \langle z,i,n |
        \psi(t) \rangle \langle \psi(t) | z,i,n' \rangle .
    \end{equation}

    The photon distribution $P_n=\rho_{nn}$ follows immediately
    from Eq.~({\ref{rhot})~:
    \begin{equation}
        \label{Pn}
        P_n(t) = \sum_{i = a,b} \int \!\! dz |\langle z,i,n |
        \psi(t) \rangle |^2 .
    \end{equation}

    The change $\delta P_n$ in the cavity photon distribution
    induced by the interaction of the cavity electromagnetic field with the incident atom is then given by
    \begin{equation}
        \delta P_n = P_n(t) - P_n(0) .
    \end{equation}

    Inserting Eqs.~(\ref{psi0}) and (\ref{psit}) into Eq.~(\ref{Pn}) and using Eqs.~(\ref{SanSbn}) and (\ref{RanRbnTanTbn}), we
    get for an incident atom of momentum $\hbar k$~:
    \begin{equation}
        \label{deltaPn}
        \delta P_n = \left \{
            \begin{array}{ll}
                 \Delta_n -\Delta_{n-m} & (n \geq m) , \\
                 \Delta_n & (n < m) .
            \end{array} \right.
    \end{equation}

    We see that if the initial state is $|a,n\rangle$ we have
    \begin{equation}
        \delta \sigma_{aa} + \delta P_{n + m} = 0 ,
    \end{equation}
    which gives an intuitive population conservation condition.

    \subsection{Reflection and transmission probabilities}

    The reflection and transmission probabilities of the incident
    atom upon the cavity are respectively given by
    \begin{mathletters}
    \label{defRandT}
    \begin{eqnarray}
        R & = & \sum_{i = a,b} \sum_{n} \int_{-\infty}^{0} \!\! dz |\langle z,i,n |
        \psi(t) \rangle |^2 , \label{defR} \\
        T & = & \sum_{i = a,b} \sum_{n} \int_{L}^{\infty} \!\! dz |\langle z,i,n |
        \psi(t) \rangle |^2 . \label{defT}
    \end{eqnarray}
    \end{mathletters}

    Inserting Eq.~(\ref{psit}) into Eqs.~(\ref{defRandT}), we get for an incident atom of momentum $\hbar k$~:
    \begin{mathletters}
    \label{RandTExpr}
    \begin{eqnarray}
        R & = & \sum_{n=0}^{\infty} w_n^2 \Big( \cos^2(\theta_n /2) |r_n^+|^2 + \sin^2(\theta_n /2) |r_n^-|^2 \Big) ,\\
        T & = & \sum_{n=0}^{\infty} w_n^2 \Big( \cos^2(\theta_n /2) |t_n^+|^2 + \sin^2(\theta_n /2)
        |t_n^-|^2 \Big) \nonumber \\
                & & + \sum_{n=1}^{m} w_{-n}^2 .
    \end{eqnarray}
    \end{mathletters}

    One verifies immediately that the results of Meyer
    \etal \cite{Mey97} about the reflection and transmission
    probabilities are well recovered by Eqs.~(\ref{RandTExpr}) when their initial conditions
    are considered. Indeed, when the atom-field system is initially in the state $|a,n\rangle$, Eqs.~(\ref{RandTExpr}) become
    \begin{mathletters}
    \label{RandTExprSystInAn}
    \begin{eqnarray}
        R & = & \frac{1}{2} \big(|r^+_n|^2 + |r^-_n|^2 \big) ,\\
        T & = & \frac{1}{2} \big(|t^+_n|^2 + |t^-_n|^2 \big) .
    \end{eqnarray}
    \end{mathletters}

    We get the same results if the atom-field system is initially in the state $|b,n\rangle$ with $n \geq
    m$, except that $n$ must be replaced by $n - m$ in
    Eqs.~(\ref{RandTExprSystInAn}). In the case $n < m$, we have
    obviously $T = 1$.

    \subsection{Final remarks}

    All the results here above (about the atomic populations, the photon statistics, and the reflection and transmission probabilities) may be very easily
    generalized for any momentum wavefunction $A(k)$ of the initial wave packet. The various expressions must simply be weighted by $|A(k)|^2$ and integrated
    over $k$. For instance, Eq.~(\ref{deltaSigmaAA}) becomes
    \begin{equation}
        \label{deltaSigmaAAWithWavePacket}
        \delta \sigma_{aa} = \int \!\! dk \, |A(k)|^2 \sum_{n=0}^{\infty} \Delta_n ,
    \end{equation}
    where $\Delta_n$ depends on $k$ through the reflection and transmission
    coefficients, $r_n^{\pm}(k)$ and $t_n^{\pm}(k)$ respectively, in $K_n$ (see Eq.~(\ref{deltaN})).

    The expressions obtained for all these various physical quantities
    are very simple in the framework of the dressed-state formalism, even though they are very general.
    They take a form much more complicated when the usual coordinates
    of the atom-field system are used (the complex coefficients $c_a$, $c_b$ and
    $c(n)$ of the atom-field states written as $(c_a |a\rangle
    + c_b |b\rangle) \otimes \sum_n c(n)|n\rangle$). Also entangled initial states may be considered by this formalism.
    The great advantage of the dressed-state coordinates was already pointed out by Jonathan \etal \cite{Jon99} who used them to express various physical quantities in the
    Jaynes-Cummings model.

    \section{Collapse-revivals}
    \label{CollapseRevivalsSection}

    Expressions (\ref{deltaSigmaAA}) and (\ref{deltaPn}) show that the features of the changes in the atomic
    populations and in the photon distribution $P_n$ with respect to the
    interaction length $\kappa L$ are directly related to the characteristics of
    $\Delta_n$. This in turn depends on the atom-field initial state (through the dressed-state coordinates)
    and on the cavity field mode profile $u(z)$ which affect the
    reflection and transmission coefficients, $r_n^{\pm}$ and
    $t_n^{\pm}$ respectively, and thus $K_n$.

    If $K_n$ have a strong oscillatory behaviour with respect to $\kappa L$, we may
    expect collapse-revivals in the population changes when several modes of the field are initially
    filled. In the following we will restrict ourselves to the
    description of the collapse-revivals when the atom is prepared
    completely in the upper or in the lower state, and the field
    is in the state $\sum_n c(n) |n\rangle$.

    When the atom is initially in the upper state $|a\rangle$, $\sigma_{bb}(t) = 1 - \sigma_{aa}(t)$
    represents the probability that a photon be emitted by the atom due to its
    interaction with the cavity. From Eq.~(\ref{sigmaAAt}), one gets that this induced
    emission probability is given by
    \begin{equation}
        \label{Pem}
        P_{em} = \sum_{n=0}^{\infty} p(n) P_{em}(n) ,
    \end{equation}
    with $p(n) = |c(n)|^2$ and
    \begin{equation}
        \label{PemN}
        P_{em}(n) = \frac{1 - \textrm{Re}(K_n)}{2} .
    \end{equation}

    When the atom is initially in the lower state $|b\rangle$,  $\sigma_{aa}(t)$ represents the probability that a cavity photon be absorbed by the atom.
    This absorption probability is identical to the induced emission probability $P_{em}$,
    except that $p(n)$ must be replaced by $p(n+m)$ in Eq.~(\ref{Pem}).

    The induced emission probability $P_{em}$ is studied hereafter for different cavity mode profiles: mesa, sech$^2$ and gaussian modes.
    Our description is restricted to the ultracold regime (incident atoms with a momentum $\hbar k$ such that $k/\kappa \ll 1$).

    \subsection{Mesa mode}
    \label{MesaModeSection}

    In the special case where the cavity field mode profile is given
    by the mesa function
    \begin{equation}
        u(z)= \left\{  \begin{array}{ll}
                        1 & \textrm{for} \,\,\, 0 < z < L \\ 0 &
                        \textrm{elsewhere}
                    \end{array}
            \right.
    \end{equation}
    the reflection and transmission coefficients $r_n^{\pm}(k)$ and $t_n^{\pm}(k)$ respectively may be calculated
    analytically. Their expression has been given for the one-photon mazer by L\"{o}ffler \etal
    \cite{Lof97}.
    Same results are obtained for the $m$-photon mazer, except that the value of the parameter $\kappa_n$ must be changed accordingly (see Eq.~(\ref{kappanDef})).
    Inserting these results into Eq.~(\ref{PemN}), one gets when $\exp(\kappa_n L) \gg
    1$ and $(\kappa_n/2k)^2 \exp(\kappa_n L) \sin(\kappa_n L) \gg 1$ that
    \begin{equation}
        \label{PemNmesa}
        P_{em}(n) = \frac{\frac{1}{2}\left[ 1 + \frac{1}{2} \sin (2 \kappa_n L) \right]}
                         {1 + (\kappa_n/2k)^2 \sin^2(\kappa_n L)} \: .
    \end{equation}

    As pointed out by L\"{o}ffler \etal \cite{Lof97}, Eq.~(\ref{PemNmesa})
    is similar to the Airy function of the classical
    optics which gives the transmitted intensity in a Fabry-Perot
    interferometer. This by no means exhibits a strong oscillatory
    behaviour. Thus, we have no chance to obtain similar
    collapse-revivals as those in the Jaynes-Cummings model when
    several mode of the field are initially filled. This is
    illustrated in Fig.~\ref{FigurePemMesaCoherentState} for the 1, 2
    and 3-photon mazers where the cavity field is taken initially
    in a coherent state ($p(n) = e^{-\bar{n}}
    \frac{\bar{n}^n}{n!}$) with a mean photon number
    $\bar{n} = 10$. A chaotic behaviour in the curves
    $P_{em}(\kappa L)$ is clearly obtained for $k/\kappa = 0.1$.

    \subsection{Sech$^2$ mode}
    \label{Sech2ModeSection}

    When the cavity field mode profile is given by the sech$^2$ function
    \begin{equation}
        u(z) = \textrm{sech}^2(z/L)
    \end{equation}
    the reflection and transmission coefficients may also be calculated analytically.
    Their expression has been given for the one-photon mazer by L\"{o}ffler \etal \cite{Lof97}.
    Again, same results are obtained for the $m$-photon mazer, except that the value of the parameter $\kappa_n$ must be changed
    accordingly. Hence, the curves $P_{em}(n)$ with respect to
    the interaction length $\kappa_n L$ are identical for the one-photon and the $m$-photon
    mazers. Two such curves have been presented by L\"{o}ffler
    \etal \cite{Lof97} for $k/\kappa_n = 0.1$ and $k/\kappa_n = 0.01$.
    These curves present well resolved resonances that get smeared for
    large values of the interaction length.
    We have calculated for the 1, 2 and 3-photon mazers the induced emission probability
    in the case of a cavity field initially in a coherent state (with $\bar{n} = 10$).
    These results are presented on Fig.~\ref{FigurePemSech2CoherentState} 
    for $k/\kappa = 0.1$. Evidence for collapse-revivals
    is shown on these figures. They are stronger in the
    case of the 3-photon mazer.

    \subsection{Gaussian mode}
    \label{GaussianModeSection}

    For a cavity field mode profile described by the gaussian function
    \begin{equation}
        \label{gaussianProfile}
        u(z)=e^{-\frac{z^2}{2 \sigma^2}}
    \end{equation}
    the reflection and transmission coefficients can no more be calculated analytically.

    We proposed recently \cite{Bas99} a numerical method for computing
    efficiently these coefficients and the induced emission
    probability $P_{em}(n)$. We compare on Fig.~\ref{FigureGaussianPemN}
    the results obtained for this probability to those calculated
    in the case of the sech$^2$ mode profile. We have considered $k/\kappa_n = 0.1$ and interaction
    lengths $\kappa_n L$ varying between 0 and 20.
    The parameter $\sigma$ in Eq.~(\ref{gaussianProfile}) was fixed to $\sqrt{2/ \pi} L$ in order to
    adopt the same normalization factor for the two profiles (identical area under the
    modes). As we pointed out in \cite{Bas99}, the
    resonances in the curves get smeared with increasing values of
    $\kappa_nL$ for both profiles. But this fact is not so marked in the case of
    the gaussian profile where the resonances still exist for longer
    interaction lengths. It is a result to be expected as the
    gaussian profile is growing more abruptly than the
    $\textrm{sech}^2$ one. Thus it is in some sense ``closer'' to the
    mesa mode, which exhibits resonances at infinity.

    We have then calculated the induced emission probability $P_{em}$ with respect to $\kappa L$ for a
    field initially in a coherent state (with $\bar{n} = 10$).
    The result is presented on Fig.~\ref{FigureGaussianCoherentState} for the 1, 2 and
    3-photon mazers.
    As the curves for the probability $P_{em}(n)$ are qualitatively similar
    in the cases of the gaussian and the sech$^2$ modes, it is not
    a surprising result that the collapse-revivals are also
    similar in both cases. Nevertheless, they are stronger for the gaussian
    potential because the oscillations in $P_{em}(n)$ are stronger too.

    Squeezing the field inside the cavity has an effect on the
    collapse-revival patterns. We have considered initial photon distributions $p(n)$ inside the
    cavity corresponding to various squeezed coherent states $|\alpha, r e^{i\theta}\rangle$, namely
    \begin{eqnarray}
        p(n) & = & \frac{(\tanh r)^n}{2^n n! \cosh r}
                   \left| H_n \left( \frac{\alpha e^{-i \theta/2}}{\sqrt{2 \cosh r \sinh r}} \right)\right|^2 \nonumber \\
             &   & \times \exp \left[ -|\alpha|^2 + \frac{1}{2} (
             e^{-i \theta} \alpha^2 + e^{i \theta} (\alpha^*)^2)
             \tanh r \right] ,
    \end{eqnarray}
    where $H_n(z)$ designates the $n^{th}$ order Hermite
    polynomial. We noticed that squeeze parameters $r$ of the order of 0.3 enhance significantly the
    collapse-revivals presented on Figs.~\ref{FigurePemSech2CoherentState} and
    \ref{FigureGaussianCoherentState} (keeping the same coherent parameter $|\alpha|^2 = 10$ and taking $\theta = 0$), while higher squeeze
    parameters tend to destroy them.

    \section{Population Trapping}
    \label{PopulationTrappingSection}

    When the atom-field initial state is such that $\sin(\theta_n)=0$, we
    get from Eq.~(\ref{deltaN}) $\Delta_n = 0$, whatever the value of $K_n$.
    In this case, we have
    \begin{equation}
        \delta \sigma_{aa} = \delta \sigma_{bb} = \delta P_n = 0 ,
    \end{equation}
    indicating that the interaction of the atom with the cavity radiation field
    has no effect on the atomic populations $\sigma_{ii}$ ($i = a,b$) and on the cavity photon
    distribution $P_n$, whatever the cavity field mode
    function, whatever the cavity interaction length $\kappa L$ and whatever the atomic initial velocity.
    We conclude that the mazer give rise to the
    perfect population trapping phenomenon, when considering zero
    temperature and no dissipation in the high-$Q$ cavity. This property holds for the ultracold, intermediate and
    thermal-atom regimes, as it is completely independent on the
    external atomic degree of freedom. For the same reason,
    it holds for any momentum wavefunction $A(k)$ of the initial
    wave packet.

    The class of states verifying $\sin(\theta_n)=0$, named \emph{perfect trapping states}, are given by
    \begin{equation}
        | \gamma^{\pm} \rangle = \frac{ \gamma^m |a\rangle \pm |b\rangle }{\sqrt{1 + |\gamma|^{2m}}}
                        \otimes \sqrt{1 - |\gamma|^2}
                        \sum_{n=0}^{\infty} \gamma^n |n\rangle ,
    \end{equation}
    where $\gamma$ is a complex number with $|\gamma| < 1$.

    Indeed, rewriting these states in terms of the dressed-state basis, we
    find
    \begin{equation}
        \label{gamma}
        | \gamma ^{\pm} \rangle = \sqrt{\frac{1 - |\gamma|^2}{1 + |\gamma|^{2m}}}
                        \left( \sum_{n=0}^{\infty} \sqrt{2} \gamma^{n+m}
                        |\pm,n\rangle \pm \sum_{n=0}^{m-1}
                        \gamma^{n} |b,n\rangle \right) .
    \end{equation}

    For each $n$ there is only a single dressed-state
    present in the sum of expression (\ref{gamma}). Depending on whether it is $|+,n\rangle$ or $|-,n\rangle$,
    we have respectively $\sin(\theta_n/2) = 0$ or $\cos(\theta_n/2) =
    0$, and so $\sin(\theta_n) = 0$ in any case.

    This give rise to another very interesting feature of the
    perfect trapping states. The reflection and transmission
    probabilities (\ref{RandTExpr}) become
    \begin{mathletters}
    \begin{eqnarray}
        R & = & \sum_{n=0}^{\infty} w_n^2 |r_n^{\pm}|^2 ,\label{RTrappingState}\\
        T & = & \sum_{n=0}^{\infty} w_n^2 |t_n^{\pm}|^2 + \sum_{n=1}^{m}
        w_{-n}^2 ,
    \end{eqnarray}
    \end{mathletters}
    with
    \begin{mathletters}
    \begin{eqnarray}
        w_{n}  & = & \sqrt{\frac{1 - |\gamma|^2}{1 + |\gamma|^{2m}}} \sqrt{2} |\gamma|^{n+m} , \\
        w_{-n} & = & \sqrt{\frac{1 - |\gamma|^2}{1 + |\gamma|^{2m}}}
        |\gamma|^{m-n} .
    \end{eqnarray}
    \end{mathletters}

    The particle moving along the $z$-axis is only sensitive to
    either a superposition of the potentials $V_n^+(z)$ or a superposition of $V_n^-(z)$, but never to both. In principle, it would be possible to imagine
    an experimental set-up where the
    particles would encounter only an effective potential well,
    instead of an effective potential hill.

    It is important to emphasize that the perfect trapping states
    do not make the cavity transparent to the incident atoms, because the
    reflection coefficient $R$ is not nullified.

    \section{Summary}
    \label{SummarySection}

    In this paper, we have studied collapse-revival patterns that appear in the changes of the atomic
    populations induced by the interaction of ultracold two-level atoms
    with electromagnetic cavities of various interaction lengths that are in resonance with an $m$-photon
    transition of the atoms. In particular, the sech$^2$ and gaussian cavity mode profiles have been considered and differences in
    the collapse-revival patterns are reported. They are stronger
    in the case of the gaussian potential.
    With the aim of such studies in view, we have
    written the quantum theory of the $m$-photon mazer by use of
    the dressed-state coordinate formalism. Simple expressions for the atomic populations, the cavity photon statistics, and the reflection and transmission
    probabilities have been given for any initial pure state of the atom-field system.
    The evidence for the population trapping phenomenon has then
    been very easily given. The trapping states written in
    Sec.~\ref{PopulationTrappingSection} have the property to
    leave, after the atom-field interaction, the cavity field and the internal atomic degrees of
    freedom at their initial value,
    independently of the cavity field mode, the cavity interaction length, and the initial atomic
    velocity.

    \acknowledgements This work has been supported by the Belgian
    Institut Interuniversitaire des Sciences Nucl\'eaires (IISN) and
    the Brazilian Conselho Nacional de Desenvolvimento Cient\'{\i}fico
    (CNPq). E. S. wants to thank for hospitality to Prof.\ Werner
    Vogel and co-workers at Rostock University in Germany.



\begin{thebibliography}{10}

\bibitem{Scu96}
M.~O. Scully, G.~M. Meyer, and H. Walther, Phys. Rev. Lett. {\bf
76},  4144
  (1996).

\bibitem{Mey97}
G.~M. Meyer, M.~O. Scully, and H. Walther, Phys. Rev. A {\bf 56},
4142
  (1997).

\bibitem{Lof97}
M. L\"offler, G.~M. Meyer, M. Schr\"oder, M.~O. Scully, and H.
Walther, Phys.
  Rev. A {\bf 56},  4153  (1997).

\bibitem{Sch97}
M. Schr\"oder, K. Vogel, W.~P. Schleich, M.~O. Scully, and H.
Walther, Phys.
  Rev. A {\bf 56},  4164  (1997).

\bibitem{Ret98}
J.~C. Retamal, E. Solano, and N. Zagury, Optics Comm. {\bf 154},
28  (1998).

\bibitem{Bas99}
T. Bastin and E. Solano, \textit{Numerical computation of
one-photon mazer
  resonances for arbitrary field modes}, to be published in Comp. Phys. Comm.
  (1999).

\bibitem{Zha99}
Z.-M. Zhang, Z.-Y. Lu, and L.-S. He, Phys. Rev. A {\bf 59},  808
(1999).

\bibitem{Ebe80}
J.~H. Eberly, N.~B. Narozhny, and J.~J. Sanchez-Mondragon, Phys.
Rev. Lett.
  {\bf 44},  1323  (1980).

\bibitem{Nar81}
N.~B. Narozhny, J.~J. Sanchez-Mondragon, and J.~H. Eberly, Phys.
Rev. A {\bf
  23},  236  (1981).

\bibitem{Yoo81}
H.~I. Yoo, J.~J. Sanchez-Mondragon, and J.~H. Eberly, J. Phys. A
{\bf 14},
  1383  (1981).

\bibitem{Fle93}
M. Fleischhauer and W.~P. Schleich, Phys. Rev. A {\bf 47},  4258
(1993).

\bibitem{Zah89}
K. Zaheer and M.~S. Zubairy, Phys. Rev. A {\bf 39},  2000  (1989).

\bibitem{Cir90}
J.~I. Cirac and L.~L. S\'{a}nchez-Soto, Phys. Rev. A {\bf 42},
2851  (1990).

\bibitem{Gea91}
J. Gea-Banacloche, Phys. Rev. A {\bf 44},  5913  (1991).

\bibitem{Yoo85}
H.~I. Yoo and J.~H. Eberly, Phys. Rep. {\bf 118},  239  (1985).

\bibitem{Cir91}
J.~I. Cirac and L.~L. S\'{a}nchez-Soto, Phys. Rev. A {\bf 44},
3317  (1991).

\bibitem{Fil86}
P. Filipowicz, J. Javanainen, and P. Meystre, J. Opt. Soc. Am. B
{\bf 3},  906
  (1986).

\bibitem{Wei99}
M. Weidinger, B.~T.~H. Varcoe, R. Heerlein, and H. Walther, Phys.
Rev. Lett.
  {\bf 82},  3795  (1999).

\bibitem{Jon99}
D. Jonathan, K. Furuya, and A. Vidiella-Barranco,
\textit{Dressed-State
  Approach to Population Trapping in the Jaynes-Cummings Model},
  quant-ph/9904067  (1999).

\end{thebibliography}


   \vspace*{3cm}

    \begin{figure}
    \begin{center}
    \noindent\fbox{\includegraphics[width=8cm, trim= 30 400 40 120,
    keepaspectratio]{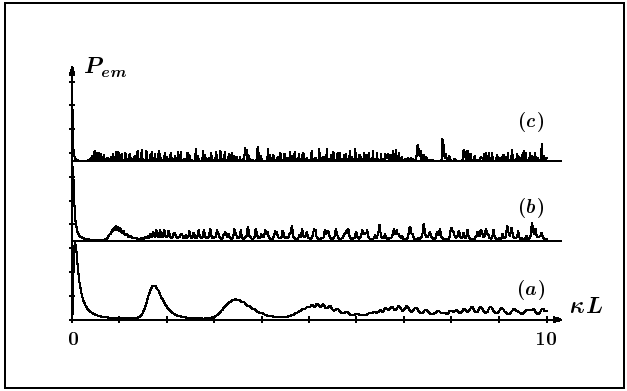}}
    \end{center}
    \caption{The induced emission probability $P_{em}$ as a function
    of the interaction length $\kappa L$ for the mesa
    mode profile and a cavity field initially in a coherent state ($\bar{n}=10$, $k/\kappa=0.1$).
    (a) 1-photon mazer, (b) 2-photon mazer, (c) 3-photon mazer. The y-scale is between 0 and 0.5 for each curve.}
    \label{FigurePemMesaCoherentState}
    \end{figure}

    \begin{figure}
    \begin{center}
    \noindent\fbox{\includegraphics[width=8cm, trim= 30 400 40 120,
    keepaspectratio]{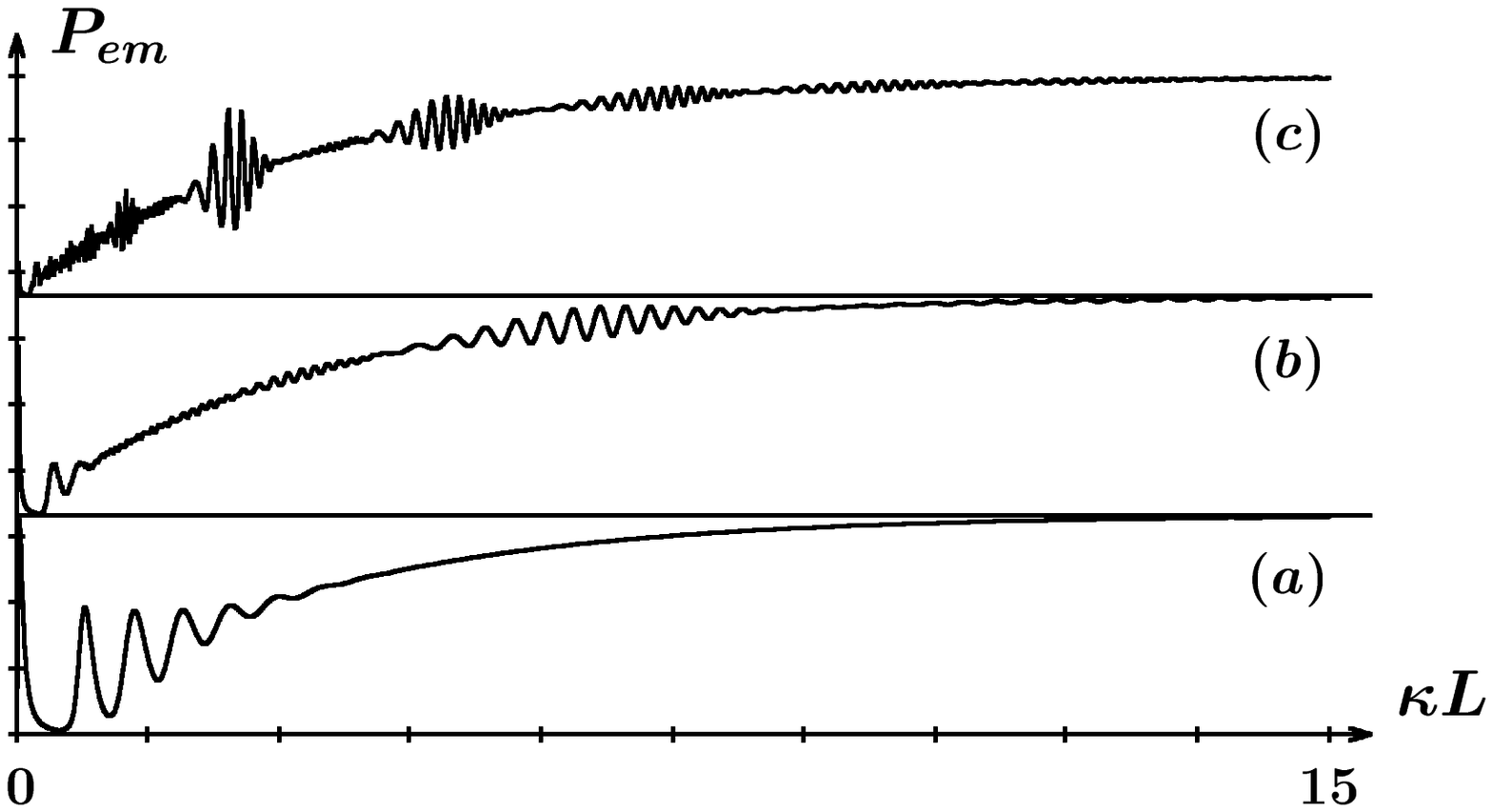}}
    \end{center}
    \caption{The induced emission probability $P_{em}$ as a function
    of the interaction length $\kappa L$ for the sech$^2$
    mode profile and a cavity field initially in a coherent state ($\bar{n}=10$, $k/\kappa=0.1$).
    (a) 1-photon mazer, (b) 2-photon mazer, (c) 3-photon mazer. The y-scale is between 0 and 0.5 for each curve.}
    \label{FigurePemSech2CoherentState}
    \end{figure}

    \begin{figure}
    \begin{center}
    \noindent\fbox{\includegraphics[width=8cm, trim= 30 400 20 120,
    keepaspectratio]{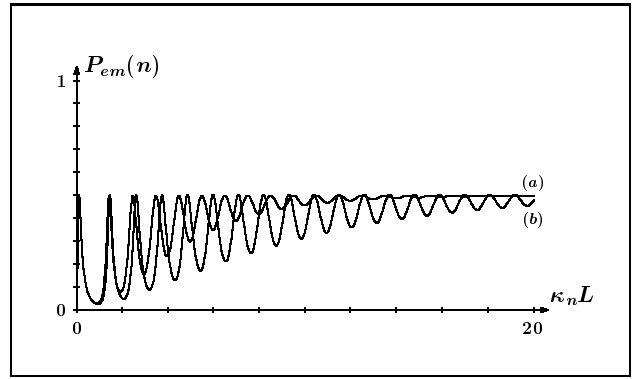}}
    \end{center}
    \caption{The induced emission probability $P_{em}(n)$ as a function
    of the interaction length $\kappa_nL$ for the $\textrm{sech}^2$
    mode profile (a) and the gaussian profile (b) ($k/\kappa_n=0.1$).}
    \label{FigureGaussianPemN}
    \end{figure}

    \begin{figure}
    \begin{center}
    \noindent\fbox{\includegraphics[width=8cm, trim= 30 400 40 120,
    keepaspectratio]{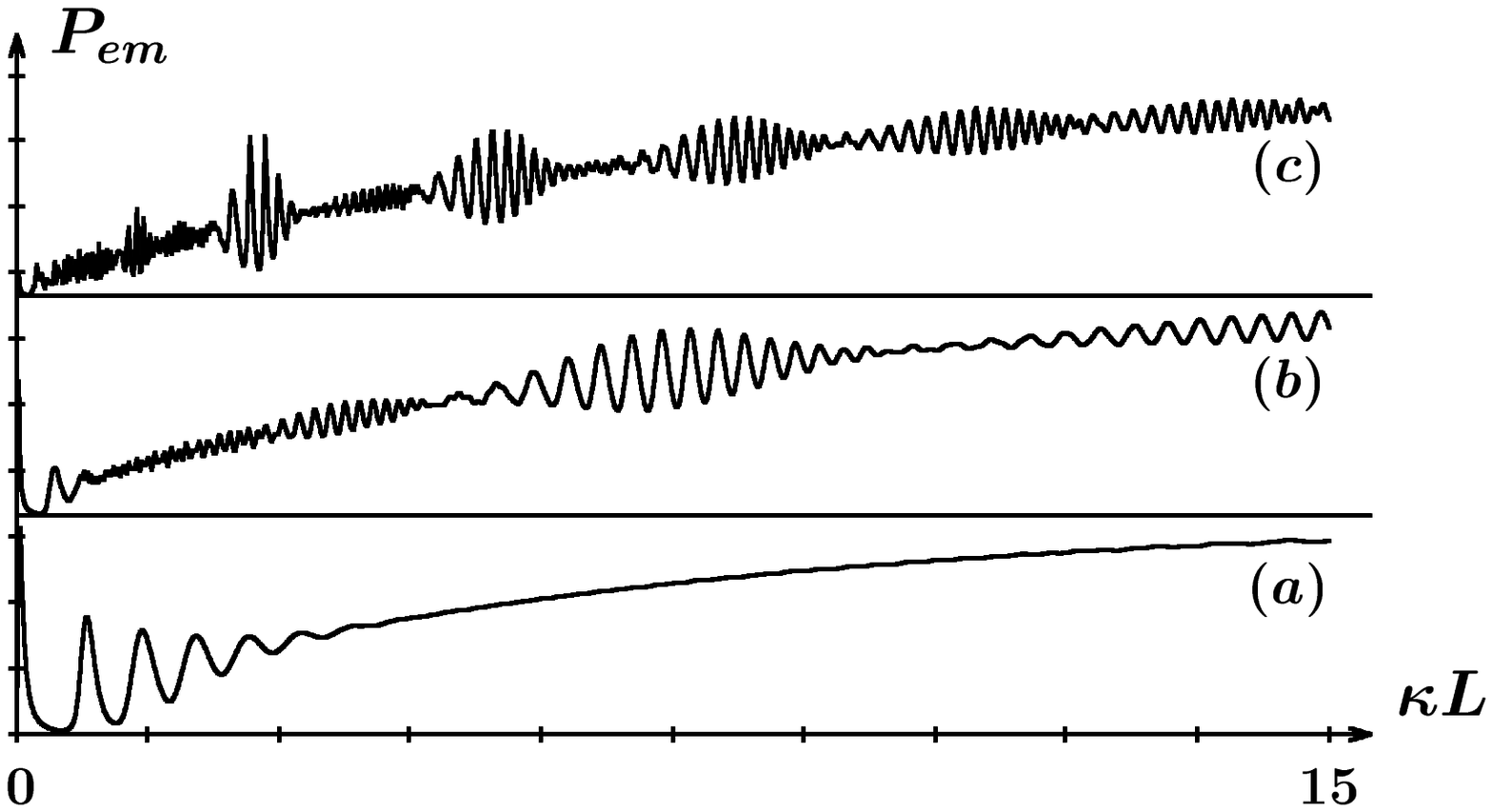}}
    \end{center}
    \caption{The induced emission probability $P_{em}$ as a function
    of the interaction length $\kappa L$ for the gaussian
    mode profile and a cavity field initially in a coherent state ($\bar{n}=10$, $k/\kappa=0.1$).
    (a) 1-photon mazer, (b) 2-photon mazer, (c) 3-photon mazer. The y-scale is between 0 and 0.5 for each curve.}
    \label{FigureGaussianCoherentState}
    \end{figure}

\end{document}